\documentclass[sigconf]{acmart}
\AtBeginDocument{%
  }

\author{YUKUN ZHANG}
\affiliation{%
  \institution{The Chinese University Of Hongkong}
  \city{HongKong}
  \country{China}}
\email{215010026@link.cuhk.edu.cn}

    \author{QI DONG}
\affiliation{%
  \institution{Fudan University}
  \city{ShangHai}
  \country{China}}
\email{19210980065@fudan.edu.cn}

\begin{document}

\title{Exploring the Head Effect in Live Streaming Platforms: A Two-Sided Market and Welfare Analysis}



\begin{abstract}
We develop a comprehensive theoretical framework to analyze live streaming platforms as two-sided markets, focusing on the “head effect” where a small subset of elite streamers disproportionately attracts viewer attention. By constructing both static and dynamic models, we capture the interplay between network effects, content quality investments, and platform policies—such as commission structures and traffic allocation algorithms—that drive traffic concentration. Our welfare analysis demonstrates that although short-term consumer utility may benefit from concentrated viewership, long-term content diversity and overall social welfare are adversely impacted. Extensive  simulations further validate our models and show that targeted policy interventions can rebalance viewer distribution and mitigate winner-takes-all dynamics. These findings offer actionable insights for platform designers and regulators in the digital economy.
\end{abstract}





\maketitle

\section{Introduction}
In recent years, the rapid emergence of live streaming platforms such as Twitch, YouTube Live, and Douyin has significantly reshaped digital economies. These platforms operate as two-sided markets by connecting content creators (i.e., streamers) with large audiences, thereby redefining content production, consumption, and monetization. On one side, streamers pursue enhanced visibility and revenue opportunities; on the other, viewers seek engaging, personalized content. This dual-sided interaction, amplified by strong network effects, algorithm-driven recommendations, and targeted platform policies, creates a dynamic environment in which market outcomes can shift rapidly.

A particularly striking phenomenon observed on these platforms is the so-called \emph{head effect} (or \emph{winner-takes-all} outcome), where a small number of top streamers capture the majority of viewership. While such concentration can boost consumer experience—by leveraging network effects that reinforce popularity—it also raises critical concerns regarding market concentration, diminished competition, and reduced content diversity. For instance, when only a handful of streamers dominate, the diversity of content and innovation from emerging creators may suffer, ultimately impacting long-term user satisfaction and the overall health of the platform ecosystem.

Despite extensive studies on two-sided markets and network effects, existing literature often overlooks the intricate interplay between algorithmic promotion, traffic allocation, and their welfare implications on live streaming platforms. In this paper, we bridge this gap by developing a comprehensive theoretical framework that not only models the traffic distribution dynamics induced by network effects and recommendation algorithms but also rigorously analyzes the economic consequences on social welfare—including consumer surplus, producer surplus, and platform profit. Our analysis reveals that platform policies, especially those governing traffic allocation and content promotion, play a crucial role in balancing the interests of both sides of the market and mitigating adverse concentration effects.

\section{Literature Overview}
The study of live-streaming platforms from the perspective of two-sided markets and the exploration of the “head effect” are relatively new areas within the field of digital economics. This literature review aims to synthesize the existing research relevant to the analysis of live-streaming platforms, network effects in two-sided markets, the head effect, and their implications for welfare, thereby laying a solid foundation for the theoretical framework and analysis presented in this paper.

\subsection{Two-Sided Markets and Network Effects}

In their seminal work, Rochet and Tirole (2003) defined two-sided markets, highlighting the significance of cross-group network effects, which are central to the functioning of platforms \cite{rochet2006two}. Evans (2002)  applied this to digital platforms, emphasizing low marginal costs and network effects. In live - streaming, an increase in the number of streamers attracts more viewers and vice versa  \cite{evans2002}.  Bartels et al. (2022) introduced innovative strategies for harnessing network effects in two-sided markets, informing platform business models aimed at balanced growth  \cite{Bartels2022}. Li et al. (2023) examined how network externalities influence channel strategy decisions in digital platforms, providing a framework analogous to the strategic decisions faced by live-streaming platforms \cite{li2023}.  Furthermore, Veljanovski (2007) outlined the dynamic nature of network effects and their impact on market structure and pricing, contributing to phenomena like the "head effect" observed in live-streaming  \cite{veljanovski2007}.

Economides (1996) corroborated these findings by demonstrating how network effects inherently drive market concentration \cite{economides1996},  while Arthur's (1989) notion of path dependence illustrated the inevitability of such market trajectories  \cite{arthur1989}. Yan (2024) extended these insights into the realm of e-commerce live streaming, positing that top anchors can more effectively foster consumer trust and loyalty compared to lesser-known streamers, thereby shaping consumer behaviors \cite{yan2024}.

\subsection{The "Head Effect" in Live - Streaming Platforms}

In the vibrant landscape of live-streaming platforms, the "head effect" has emerged as a central phenomenon, attracting significant academic attention. This section reviews the existing literature to comprehensively understand its nature, causes, and consequences.


Several factors Contribute to the head effect. Firstly, platform policies play a pivotal role in shaping the head effect. Yang et al. (2022) pointed out that commission rates act as a powerful incentive mechanism. Platforms often offer lower commission rates to head streamers, which not only increases their earnings but also encourages them to remain on the platform \cite{Yang2022}. Regarding traffic allocation algorithms, Soboleva (2025) revealed that most platforms’ recommendation algorithms prefer distributing advertising traffic to top streamers to maximize revenue  \cite{soboleva2025optimaltrafficallocationmultislot}. Secondly, network effects are fundamental to the formation of the head effect. Lin et al. (2024) stated that due to the social nature of live streaming, viewers are more willing to stay in crowded live-streaming rooms. Interactions among viewers and real-time activities initiated by streamers establish a sense of social presence, further affecting the income of streamers and platforms (Lin et al.,2024) \cite{lin2024}. Finally, the unique attributes of streamers themselves are also crucial. Top anchors or idol groups usually possess outstanding professional skills. For example, well-known brand ambassadors are often invited to shopping live broadcasts to provide viewers with interactive shopping experiences.   Fatihah et al. (2024) emphasized that top anchors  significantly influence the audience  views and preferences for certain products, thereby guiding purchase decisions \cite{fatihah2024}.

The aforementioned factors contributing to the head effect have several adverse implications. Firstly, the head effect significantly impacts the market structure of live-streaming platforms.  According to Peng (2022), the high concentration of viewership among a few head streamers leads to a monopolistic-like market structure. Small and medium-sized streamers struggle to gain sufficient visibility and resources, potentially limiting innovation and competition in the long run \cite{peng2022}.  Additionally, from a consumer behavior perspective, the head effect influences viewers' choices. Head streamers plays a crucial mediating role between customer experience and consumer purchase intention, leading consumers to rely heavily on head streamers' recommendations and potentially overlook other suitable products and brands \cite{chen2023}.

\subsection{ Policy Interventions}

To address the negative welfare implications of market concentration, various policy interventions can be implemented to promote a more equitable distribution of viewers and support diverse content creation. 

 Bakshy et al. (2012) asuggested adjusting platform algorithms to ensure a fairer distribution of viewer traffic among streamers, thereby mitigating the head effect \cite{Bakshy_2012}. Zhang et al. (2023) recommended reducing commission rates, which may enable smaller streamers to retain more of their revenue and encourage a more diverse streamer ecosystem  \cite{zhang2023}. Williamson et al. (1986) emphasized the necessity of antitrust measures to prevent monopolistic practices and promote competition, ensuring that no single platform or group of platforms can dominate the market \cite{baumol1986williamson}.

While existing literature provides a comprehensive understanding of two-sided markets, network effects, and the head effect in live-streaming platforms, several gaps remain. Most studies focus on identifying and describing these phenomena without thoroughly exploring the dynamic interactions between platform policies and streamer behaviors over time. Additionally, there is limited empirical evidence on the long-term welfare implications of market concentration beyond theoretical assessments. Moreover, while various policy interventions have been proposed, there is a lack of consensus on the most effective combination of policies to balance market concentration with platform growth and innovation.

.

\section{Structure and Contributions}

The paper is structured as follows: Section 4 introduces a static model to analyze viewer-streamer interactions and network effects, highlighting how platform policies contribute to market concentration. Section 5 extends this to a dynamic model, examining the evolution of viewer behavior and streamer strategies, with a focus on path dependence and the head effect. Section 6 evaluates the welfare impacts of traffic concentration on consumers, streamers, and platforms. Section 7 presents  results, including sensitivity analyses and policy evaluations, while Section 8 concludes by summarizing key findings and proposing future research directions.

This paper integrates static and dynamic models to analyze the impact of network effects and platform policies on traffic concentration in live streaming platforms. It contributes to the literature by demonstrating how strong network effects lead to market concentration, conducting a welfare analysis that highlights the trade-offs between efficiency and market fairness, and proposing policy interventions to optimize social welfare.  Simulations further validate the theoretical models, offering practical insights for platform regulation and the mitigation of traffic concentration.

\section{Static Analysis}
\label{sec:static_analysis}

In this section, we construct a static model to deeply analyze the interactions between streamers (content providers) and viewers (users) on live-streaming platforms.

\begin{itemize}
    \item \textbf{Core Objectives}: We focus on how platform policies (such as commission rates, visibility algorithms, and traffic distribution strategies) shape market outcomes and explore the possibility of multiple equilibria triggered by network effects.
    \item \textbf{Key Assumptions}: During the analysis, we assume that the total number of viewers $M$  remains constant, and pricing decisions $p_i$ (such as subscription fees or pay-per-view charges) can be set autonomously by either the platform or the streamers.
\end{itemize}

\subsection{Model Construction}
\label{subsec:model_construction}

\noindent
\textbf{Subject Setting:}  Consider two groups connected by the platform. The set of streamers  
 $\displaystyle I = \{1, 2, \dots, N\}$ is dedicated to producing live-streaming content and aims to attract viewers and gain profits. The set of viewers $ J = \{1, 2, \dots, M\}$ consumes content and may pay tips, subscriptions, or other fees.

\vspace{1em}
\noindent
\textbf{Utility and Profit Functions:}
\begin{itemize}
    \item \textbf{Viewer Utility Function:} The utility function of viewer $j$ watching streamer $i$ is

    \begin{equation}
    \label{eq:viewer_utility}
     U_{ij} = \alpha_i q_i - p_i + \beta n_i + \epsilon_{ij}.
    \end{equation}
    where,
    \begin{itemize}
        \item $\alpha_i$ (intrinsic attractiveness) reflects the overall appeal or entertainment value of streamer $i$, which is a comprehensive manifestation of the streamer's personal traits, image, popularity, etc., and is independent of the current number of viewers and other short - term variables.
        \item $q_i$ (content quality) represents the production value, consistency, or topic relevance of streamer $i$'s content. High - quality content usually requires more resource input and professional skills.
        \item $p_i$ (price or cost) indicates the fee that viewer $j$ needs to pay to watch streamer $i$, which may vary depending on platform strategies, streamer popularity, and content types.
        \item  $\beta$ (network - effect parameter) measures the additional attractiveness of streamer $i$ to viewer $j$ based on the current number of viewers $n_i$.
    \end{itemize}
    A larger $\beta$ means stronger network effects, that is, viewers are more inclined to watch live-streams with a larger number of existing viewers. $n_i$ (audience size) is the number of viewers currently watching streamer $i$. $\epsilon_{ij}$ (random component) captures the idiosyncratic preferences of viewers and is assumed to be independently and identically distributed among viewers and streamers. It represents some individual differences that are difficult to explain with other parameters in the model, such as the viewer's mood at a specific moment or accidental interests.

    \item \textbf{Streamer Profit Function:}  The profit function of streamer $i$ is
    \begin{equation}
    \label{eq:streamer_profit}
    V_i \;=\; (1 - \tau)\, R\, n_i \;-\; c\bigl(q_i\bigr),
    \end{equation}
    where,
    \begin{itemize}
        \item $\tau$ (commission rate) is the proportion that the platform extracts from streamer $i$'s revenue, which directly affects the streamer's actual income.
        \item $R$ (average revenue per viewer) encompasses tips, advertising revenue, or subscription payments from each viewer, reflecting the average level of economic return that the streamer can obtain from each viewer.
        \item $c(q_i)$ (cost function) is strictly increasing and convex in $q_i$.
    \end{itemize}
    This means that as the content quality improves, the cost required to increase each unit of quality gradually increases, reflecting the increasing marginal cost characteristics of producing high-quality content in terms of resources, time, and manpower.
       
\end{itemize}

\subsection{Equilibrium Analysis}
\noindent
\textbf{Viewer Choice Probability:} Based on the above - mentioned utility function, we adopt the multinomial logit (MNL) framework to describe viewer $j$'s choice of streamer $i$. The probability that viewer $j$ chooses streamer $i$ is 
\begin{equation}
    P_i = \frac{\exp\left( \alpha_i q_i - p_i + \beta n_i \right)}{\sum_{k=1}^N \exp\left( \alpha_k q_k - p_k + \beta n_k \right)}.
\end{equation}
 This formula is based on the principle of utility maximization. The denominator is the sum of the exponentiated utilities of all streamers, and the numerator is the exponentiated utility of streamer $i$. It represents the relative possibility of the viewer choosing streamer $i$ among all available streamers.

\vspace{1em}
\noindent
 \textbf{Expected Number of Viewers:} The expected number of viewers for streamer $i$ is $n_i=M P_i$, that is, the total number of viewers $M$ multiplied by the probability $P_i$ that the viewer chooses streamer $i$. Substituting the expression of $P_i$ into the formula for $n_i$, we get
 \begin{equation}
     n_i = M \cdot \frac{\exp\left( \alpha_i q_i - p_i + \beta n_i \right)}{\sum_{k=1}^N \exp\left( \alpha_k q_k - p_k + \beta n_k \right)},
 \end{equation}
which is an implicit system of equations about the variables  $\{n_i,q_i\}$.
 
\vspace{1em}
\noindent
 \textbf{Existence and Uniqueness of Equilibrium:} Assume that the cost function $c(q_i)$ is strictly convex and all utility components are continuous. When the network-effect parameter \(\beta\) is below a certain threshold that avoids excessive feedback, there exists a unique equilibrium $(n_i^*, q_i^*)$. The proof process usually relies on fixed - point theorems, such as the Brouwer Fixed-Point Theorem. For a function $F(n_i, q_i)$, if it is continuous on a closed and bounded set, then there exists at least one fixed - point in this set, that is, $F(n_i^*, q_i^*)=(n_i^*, q_i^*)$, and the fixed-point here is the equilibrium point. In our model, by constructing an appropriate function and combining the restrictive conditions of $\beta$ the uniqueness of the equilibrium can be proved. When $\beta$ exceeds the critical threshold, the feedback loop will lead to multiple equilibria or highly concentrated equilibria. That is, a small initial advantage in viewership will be amplified, resulting in most users concentrating around one or several streamers, forming the "head effect".

\vspace{1em}
 The static analysis lays the foundation for understanding the basic market mechanisms of live-streaming platforms. However, it fails to consider the evolution of viewer numbers and content quality over time. In reality, streamer strategies and viewer preferences change dynamically, which may introduce path-dependence. Therefore, in the following part, we will extend the model to a dynamic framework.

\section{Dynamic Analysis}
\label{sec:dynamic_analysis}
In this section, we incorporate the time dimension $t\geq0$ based on the static model and deeply study the dynamic evolution process of viewer numbers $\{n_i(t)\}$ and content quality $\{q_i(t)\}$ under the influence of network effects and platform policies, as well as potential path-dependence and multiple-equilibrium results.

\begin{itemize}
    \item \textbf{Key Questions}: We mainly explore how initial conditions affect the long-term distribution of viewers, the conditions for the emergence of a stable steady-state and its uniqueness, whether strong network effects will lead to "winner-takes-all" or "duopoly" outcomes, and how the platform should intervene dynamically to mitigate excessive concentration.
    \item \textbf{Model Adjustments}: We retain the same set of streamers and viewers as in the static model but make a series of adjustments to capture time dynamics, such as adopting continuous time, considering time-varying attractiveness (although it is simplified to constants for now, leaving room for future extensions), a fixed viewer pool (and discussing the impact of relaxing this assumption), and no explicit switching costs.
\end{itemize}
\subsection{Dynamic Equations}
 
\noindent
 \textbf{Dynamics of Viewer Numbers:}
 \begin{equation}
     \frac{d n_i(t)}{d t} = \gamma \left[ M P_i(t) - n_i(t) \right].
 \end{equation}
Here, $\gamma > 0$ is the adjustment rate, which determines the speed at which viewers adjust their viewing choices according to expected utility. If \(n_i(t)\) is lower than its "expected" level $MP_i(t)$, it means that according to the current utility calculation, more viewers should watch streamer $i$, so $n_i(t)$ will increase; conversely, if $n_i(t)$ is higher than $MP_i(t)$, then $n_i(t)$ will decrease. $P_i(t)$ is the probability that a viewer chooses streamer $i$ at time $t$, and its form is similar to $P_i$ in the static model, that is, 
\begin{equation}
    P_i(t) = \frac{\exp\left( \alpha_i q_i(t) - p_i + \beta n_i(t) \right)}{\sum_{k=1}^N \exp\left( \alpha_k q_k(t) - p_k + \beta n_k(t) \right)}.
\end{equation}

It reflects the relative possibility of a viewer choosing streamer $i$ at time $t$ based on factors such as the streamer's attractiveness, content quality, viewing cost, and the number of viewers at that time.

\vspace{1em}
\noindent
 \textbf{Dynamics of Content Quality:}
 \begin{equation}
\frac{d q_i(t)}{dt} = \eta_i\left[\,(1 - \tau) R \frac{\partial n_i(t)}{\partial q_i(t)} - c'\bigl(q_i(t)\bigr)\,\right].
 \end{equation}
Here, $\eta_i > 0$ is the speed at which streamer $i$ adjusts the content quality, representing the streamer's ability and efficiency to respond to market feedback and change the content quality. $(1 - \tau)R\frac{\partial n_i(t)}{\partial q_i(t)}$ measures the increase in marginal revenue brought about by a small improvement in content quality. That is, a small change in content quality will cause a change in the number of viewers, and then affect the revenue. And $(1 - \tau)R$ is the actual revenue brought by each viewer, so their product is the marginal revenue. $c'(q_i(t))$ is the marginal cost of improving the content quality, which is the first - order derivative of the cost function $c(q_i(t))$ with respect to $q_i(t)$.

When the marginal revenue is greater than the marginal cost, improving the content quality will increase the profit, so $\frac{d q_i(t)}{dt}>0$ and the content quality will be improved; conversely, when the marginal cost is greater than the marginal revenue, the content quality will decrease. For $\frac{\partial n_i(t)}{\partial q_i(t)}$, in the standard logit model, by taking the derivative of \(n_i(t) = M P_i(t)\) with respect to \(q_i(t)\), we can get $\frac{\partial n_i(t)}{\partial q_i(t)} = M\alpha_iP_i(t)(1 - P_i(t))$. This is because $P_i(t)$ is a function of $q_i(t)$. Using the chain-rule for composite functions, first take the derivative of $\exp(\alpha_iq_i(t) - p_i+\beta n_i(t))$ with respect to $q_i(t)$ to get $\alpha_i\exp(\alpha_iq_i(t) - p_i+\beta n_i(t))$, and then according to the quotient-rule for derivatives, combined with the denominator $\sum_{k = 1}^N \exp(\alpha_kq_k(t) - p_k+\beta n_k(t))$, after a series of simplifications, we can obtain 
\begin{equation}
    \frac{\partial n_i(t)}{\partial q_i(t)} = M\alpha_iP_i(t)(1 - P_i(t)).
\end{equation}

\subsection{Path-Dependence and Head Effect}
\noindent
 \textbf{Path-Dependence Mechanism:} When $\beta$ is large enough, small differences in initial conditions $\{n_i(0)\}$ or $\{q_i(0)\}$ will be amplified over time. A streamer with a slightly larger initial number of viewers may attract a disproportionate number of new viewers, further increasing $n_i(t)$, which in turn encourages them to invest more in $q_i(t)$, forming a positive feedback loop.

 \vspace{1em}
\noindent
 \textbf{Impact on Market Concentration:} Under moderate $\beta$, the system may converge to a unique interior equilibrium with partial concentration. When $\beta$ is high, a "winner-takes-all" scenario may occur, where one or a few streamers dominate. Through theorems (such as "In the dynamic model, when $\beta$ is large, small advantages in $\{n_i(0)\}$ or $\{q_i(0)\}$ will lead to significantly different long-term results, potentially resulting in the head effect"), this phenomenon can be theoretically elaborated.

 \vspace{1em}
 \begin{figure}[h!] 
    \centering
    \includegraphics[width=\columnwidth]{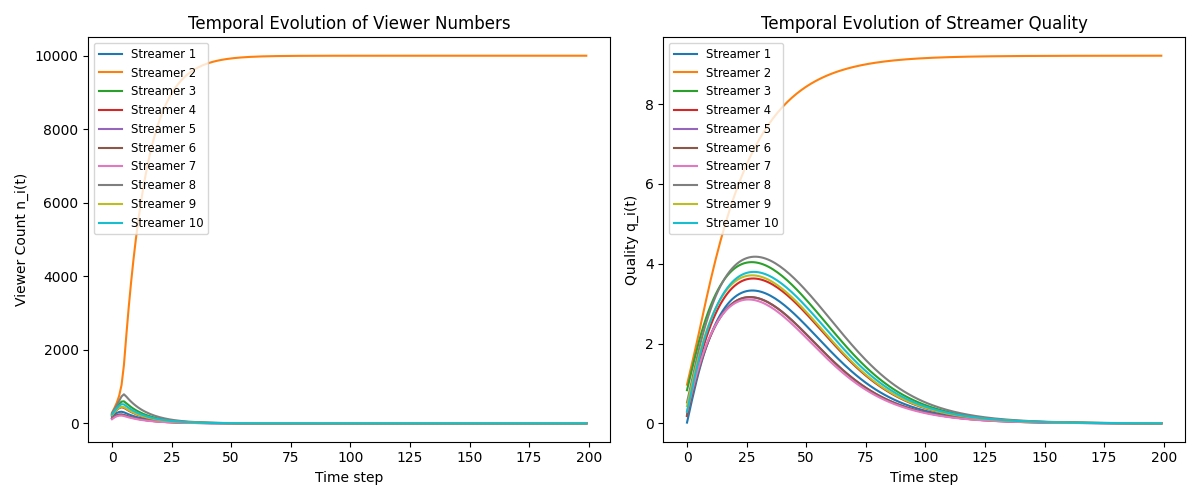}
    \caption{Temporal Evolution of Viewer Numbers and Streamer Quality.}
    \label{fig:phase_diagram}
\end{figure}
\noindent
 \textbf{Temporal Evolution of the Head Effect:} When network effects are strong, the dynamic system usually exhibits the "head effect", that is, a small number of streamers occupy a disproportionate share of viewers after a period of time. A phase diagram (such as the relationship diagram between $n_i(t)$ and $q_i(t)$ showed in Figure ~\ref{fig:phase_diagram} ) can be generated through numerical simulations to show how the state $(n_i(t), q_i(t))$ evolves towards a corner or near-corner solution, leaving other streamers with very few viewers.

 \vspace{1em}
 Compared with the static analysis, the dynamic analysis reflects the market dynamics of live-streaming platforms more realistically, shows how the system gradually reaches equilibrium and the long-term results under different conditions, and provides a more timely and targeted basis for the platform to formulate reasonable policies.

\section{Welfare Economics Analysis}
\label{sec:welfare_analysis}

Building on the dynamic model in Section~\ref{sec:dynamic_analysis}, this section explores the welfare implications of strong network effects and the "head effect" in live - streaming platforms. We analyze how viewer distribution affects consumers, producers, and the platform in the short and long term, and how platform policies can optimize overall welfare.

Based on standard welfare economics, we define:
\begin{itemize}
    \item \textbf{Consumer Surplus (CS):}  The net utility viewers gain over the price paid.
    \item \textbf{Producer Surplus (PS):} Streamers' total profit minus content production costs.
    \item \textbf{Platform Profit ($\Pi$):} Revenue from commissions and other fees.
    \item \textbf{Total Social Welfare ($W$):} $W = CS + PS+\Pi$.
\end{itemize}
We'll study how network effects, market evolution, and platform policies impact these measures, and whether the "head effect" benefits or harms overall welfare.

\subsection{Consumer Surplus (CS)}

CS measures viewers' net benefit. We model it as 
\begin{equation}
    CS=\sum_{j = 1}^M(U_j - p),
\end{equation}
where $U_j$ is viewer $j$'s utility, and $p$ is the price (which can be 0 for free-content scenarios).

When a few top streamers attract most viewers, niche-content-loving viewers may be left out. Although large-audience channels offer high utility due to network effects, the loss of content variety can reduce overall CS, especially as smaller creators leave in the long term.

Strong network effects create path-dependence. Early viewer concentration on top streamers reduces alternatives. Short-term CS may be high as viewers flock to popular channels, but long-term CS may decline due to less innovation and variety.

\subsection{Producer Surplus (PS)}

PS is calculated as 
\begin{equation}
    PS=\sum_{i = 1}^N[(1-\tau)R n_i - c(q_i)]
\end{equation}
 with $(1 - \tau)R n_i$ being streamer $i$'s revenue and $c(q_i)$ the production cost.

The head effect makes top streamers capture most revenue, while smaller streamers face low returns and may exit. Although total PS can be high with a few large streamers, lack of competition reduces incentives for quality and innovation, harming long-term content diversity and welfare.

The evolution of content quality $q_i(t)$ favors streamers with large audiences. Smaller creators may not survive, so short-term PS may be dominated by successful streamers, but future competition and innovation are at risk.

\subsection{Platform Profit ($\Pi$)}

Platform profit is 
\begin{equation}
    \Pi=\tau R M
\end{equation}
 (assuming $M$ active viewers and $\tau$ commission rate). Other revenue sources can be added in more complex models.

In the short run, concentrating viewers on top streamers cuts costs and boosts efficiency. But over time, the head effect can reduce diversity and engagement, lowering long-term $\Pi$. Balancing short-and long-term goals is crucial for platform promotion strategies.

\subsection{Optimizing Traffic Allocation and Policy Intervention}

To counter the negative impacts of excessive concentration, the platform can allocate traffic and promote content to maximize $W$. Let $\theta_i$ be the promotion share for streamer $i$.

We aim to 
\begin{equation}
    \max_{\{\theta_i\}}W(\theta_i)=CS(\theta_i)+PS(\theta_i)+\Pi(\theta_i)
\end{equation}
 subject to $\theta_i\geq0$, $\sum_{i = 1}^N\theta_i = 1$, and other feasibility constraints. In a dynamic setting, $\theta_i(t)$ can be adjusted over time.

 Promoting smaller streamers may cut short-term profit but preserve diversity for future CS. Focusing on top streamers maximizes immediate returns but weakens long - term competition. The optimal $\theta_i$ changes over time.

 Solving the optimization equation could yield a proposition about interior vs. corner solutions for $\theta_i$, guiding how to support smaller creators relative to popular ones. The marginal gain in $W$ from changing $\theta_i$ depends on its impact on CS, PS, and $\Pi$.
  \begin{figure}[h!] 
    \centering
    \includegraphics[width=\columnwidth]{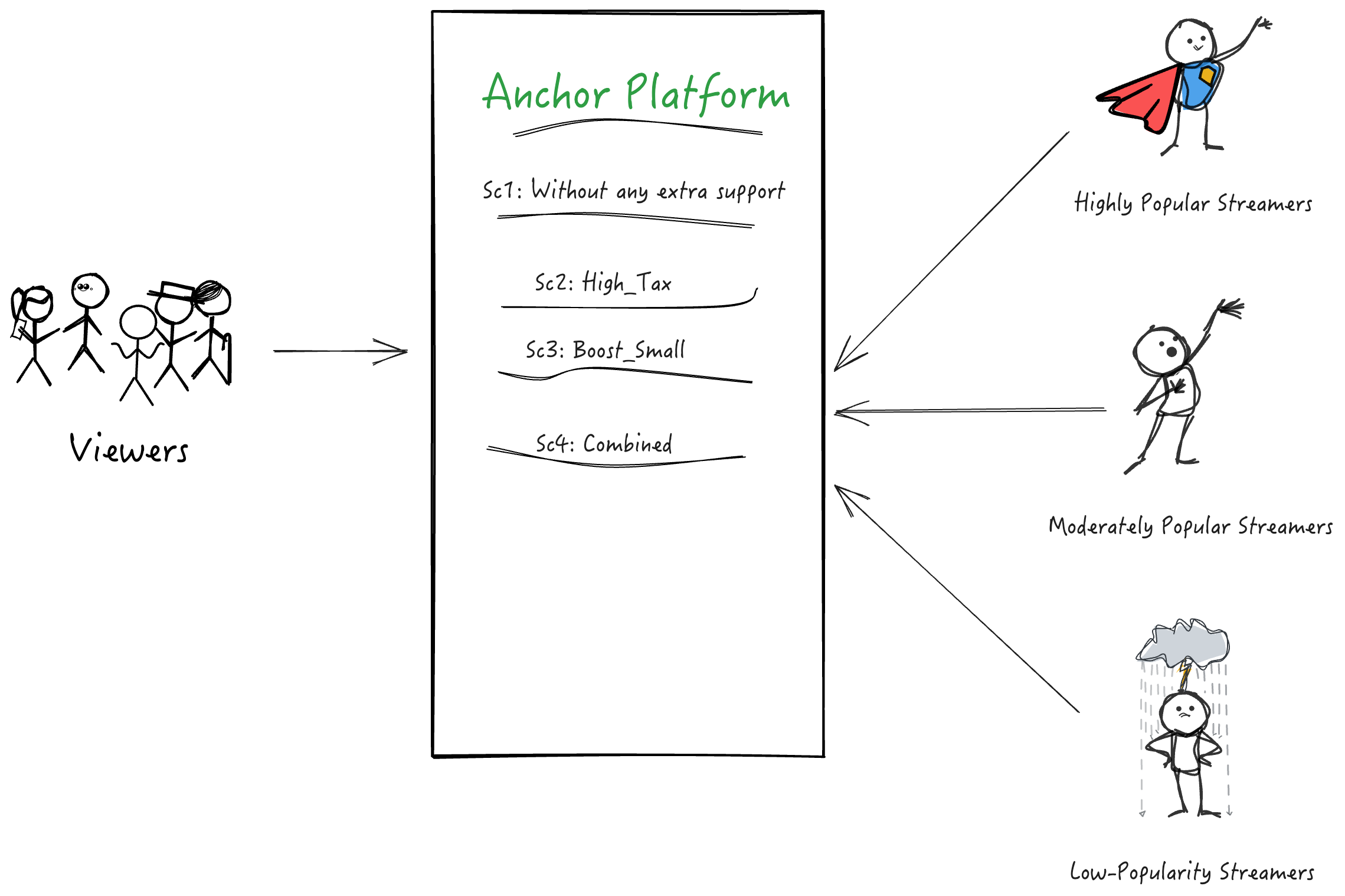}
    \caption{Interaction between streamers and viewers.}
    \label{fig:streaming}
\end{figure}

 \subsection{Conclusion}
 The "head effect" creates a complex relationship among CS, PS, and $\Pi$. While network effects can boost short-term consumer utility, they can erode long-term welfare due to less diversity. The platform may gain short-term efficiency but risk losing engagement.
 
Excessive concentration can lock the market into an unproductive equilibrium. A well-planned traffic allocation policy, adaptable over time, is needed to balance platform profit and content ecosystem health. Future research could include multi-platform competition, viewer differences, and more complex revenue models to better reflect real-world live-streaming.


\section{Experimental evaluation}
We use real user data to validate our strategy, and the theoretical model of live-streaming platforms. The main objectives of the experiments are:

\begin{table*}[htbp]
\centering
\caption{Experimental Metrics for Different Scenarios}
\label{tab:exp_metrics}
\begin{tabular}{lcccccc} \hline
Scenario     & Gini Coefficient & Top 3 Share & Viewer Mobility & Tail Share & Avg. Satisfaction & Quality Improvement \\ \hline
Baseline     & 0.5557         & 0.4590       & 6.6884        & 0.5410      & 0.9893           & -0.1344           \\
High\_Tax    & 0.4816         & 0.4000       & 9.5483        & 0.6000      & 0.7966           & -0.0230           \\
Boost\_Small & 0.3188         & 0.3190       & 12.1660       & 0.6810      & 2.1232           & -0.0356           \\
Combined     & 0.2636         & 0.3010       & 18.8245       & 0.6990      & 1.7533           & 0.0966            \\ \hline
\end{tabular}
\end{table*}

\begin{enumerate}
    \item To verify the spontaneous emergence of the head effect in the experiment.
    \item To assess the impact of platform intervention policies (e.g., increasing exposure for small and medium streamers, raising commission rates for top streamers) on traffic concentration and social welfare.
    \item To observe viewers' integrated responses to streaming content quality, streamer popularity, and platform recommendations, and check their consistency with theoretical assumptions.
\end{enumerate}


\subsection{Experimental Framework}

\label{sec:framework}
We categorize streamers into high, medium, and low-popularity groups and audiences into different interaction-willingness groups. The simulation comprises three phases:

\begin{enumerate}
    \item \textbf{Baseline Testing:} Without any extra support or policy intervention, streamers are ranked by viewer count to observe the head effect under pure network effects.
    
    \item \textbf{Single Policy Intervention:}  Individual policies (e.g., taxing top streamers, boosting small streamers, subsidizing small streamers) are tested separately, and the results are compared with the baseline.
    
    
    \item \textbf{Combined Policy Intervention:} 
    Multiple policies are implemented simultaneously to evaluate their cumulative effects on traffic, concentration, and social welfare.
\end{enumerate}

\subsection{Simulation Result}
We did an A/B Test based on the grouping in Section 
 \ref{sec:framework}
\subsubsection{Baseline scenario}
\begin{figure}[htbp]
    \centering
    \includegraphics[width=\columnwidth]{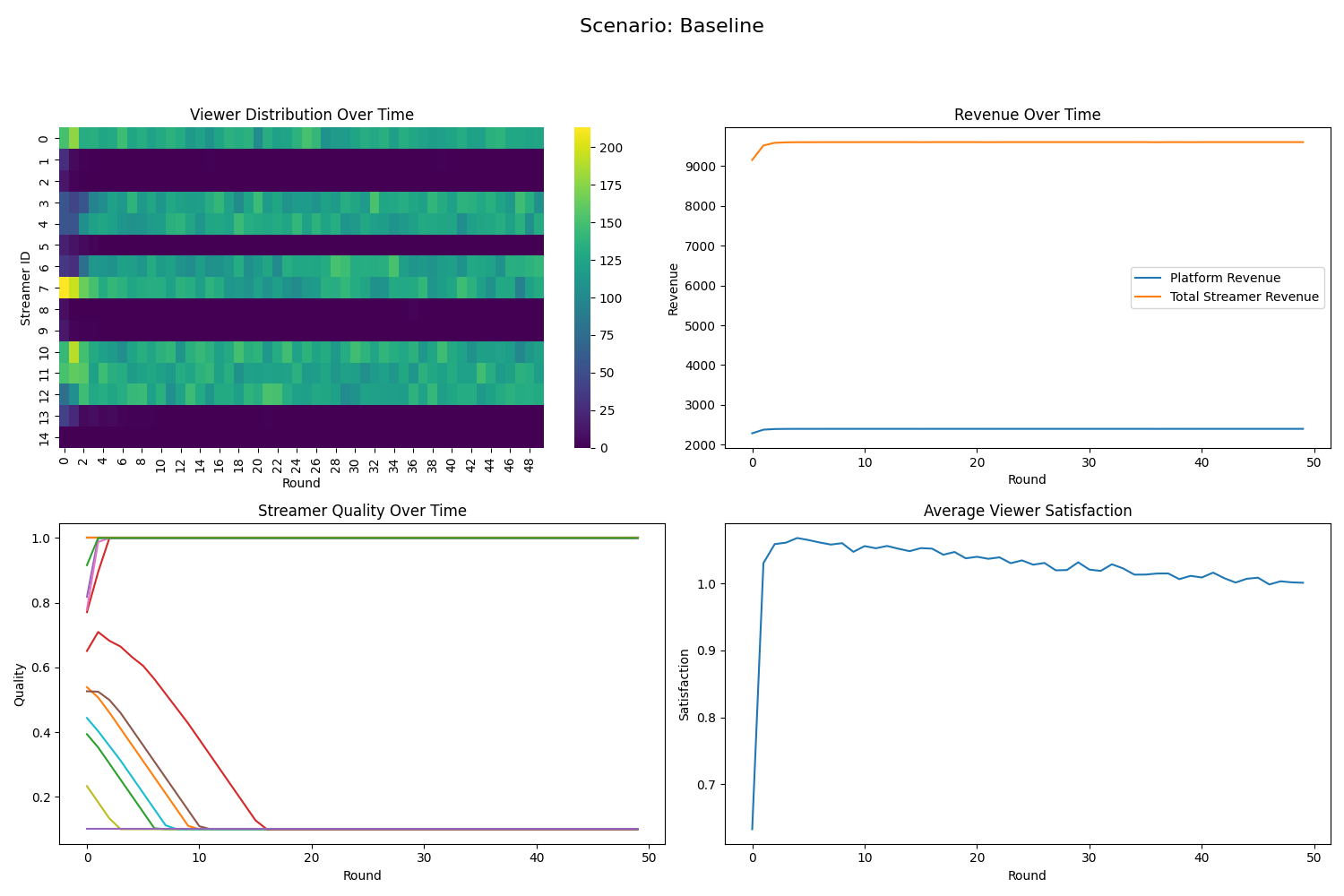}
    \caption{Baseline scenario: viewer distribution, revenues, streamer quality, and average satisfaction.}
    \label{fig:figure1}
\end{figure}

The \emph{Baseline} results in Figure 2 support our theoretical model:

\begin{itemize}
    \item \textbf{Viewer Distribution :}  
    A few streamers gradually attract more viewers, indicating the network effect. The moderate overall concentration is reflected by a Gini coefficient of about 0.556.
    \item \textbf{Revenue Trends:}  
   Total streamer revenue is initially high but stabilizes as viewer preferences converge. Platform revenue tracks a fraction of streamer earnings.
    \item \textbf{Streamer Quality Over Time:}  
    Most streamers invest in quality, yet only some maintain high levels. Diminishing returns on quality investment for certain streamers are consistent with our cost-function assumptions.
    \item \textbf{Average Viewer Satisfaction:}  
    Satisfaction rises rapidly as viewers find appealing channels and levels off near 1.0, confirming the role of network effects in viewer retention.
\end{itemize}

This scenario validates the theoretical framework, showing a self-reinforcing popularity mechanism with moderate diversity.


\subsubsection{Policy Intervention Analysis}
\label{sec:analysis_scenarios}

\begin{figure}[htbp]
    \centering
    \includegraphics[width=\columnwidth]{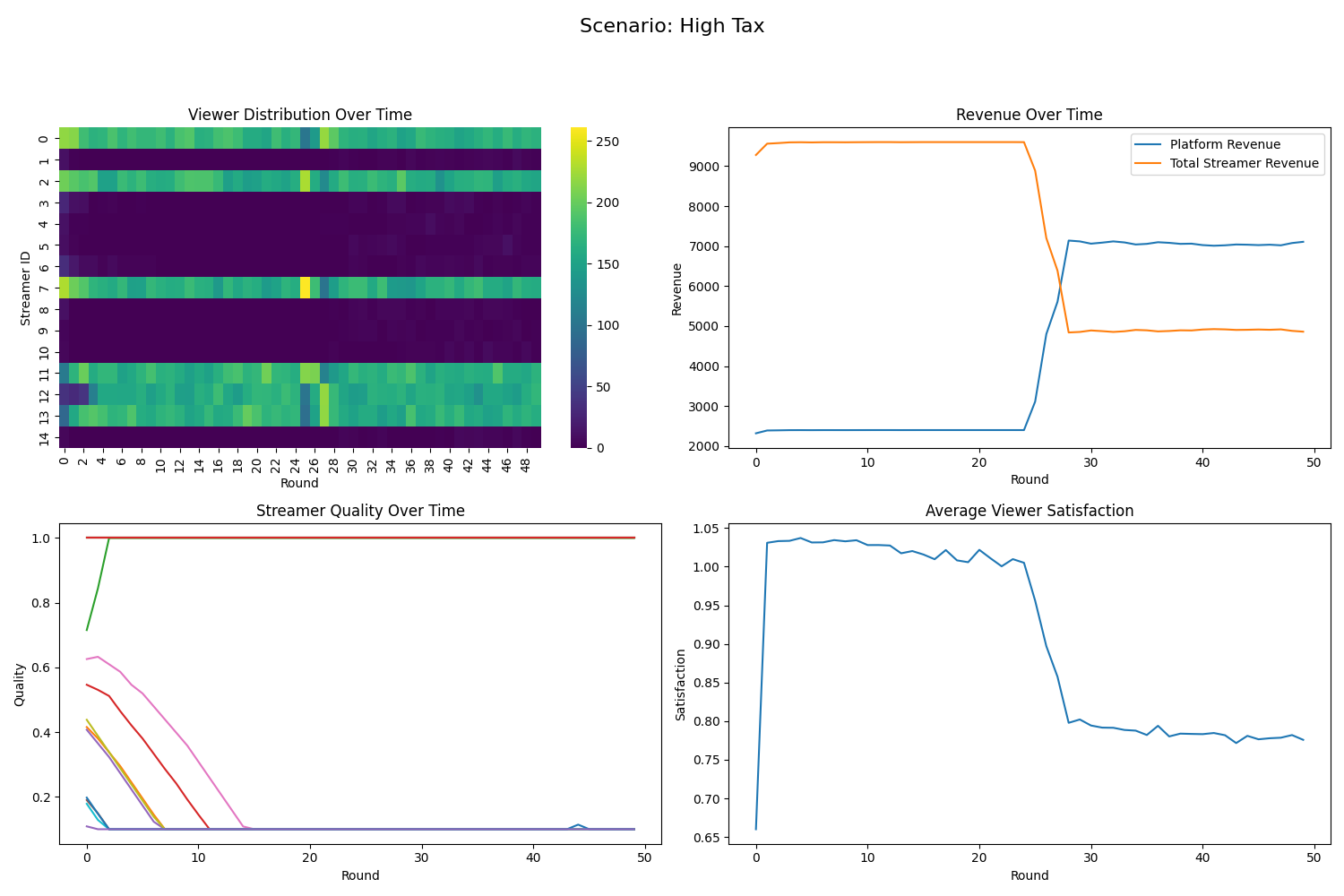}
    \caption{High\_Tax scenario: Viewer distribution, revenue trends, streamer quality, and average satisfaction over time.}
    \label{fig:figure2}
\end{figure}

\begin{figure}[htbp]
    \centering
    \includegraphics[width=\columnwidth]{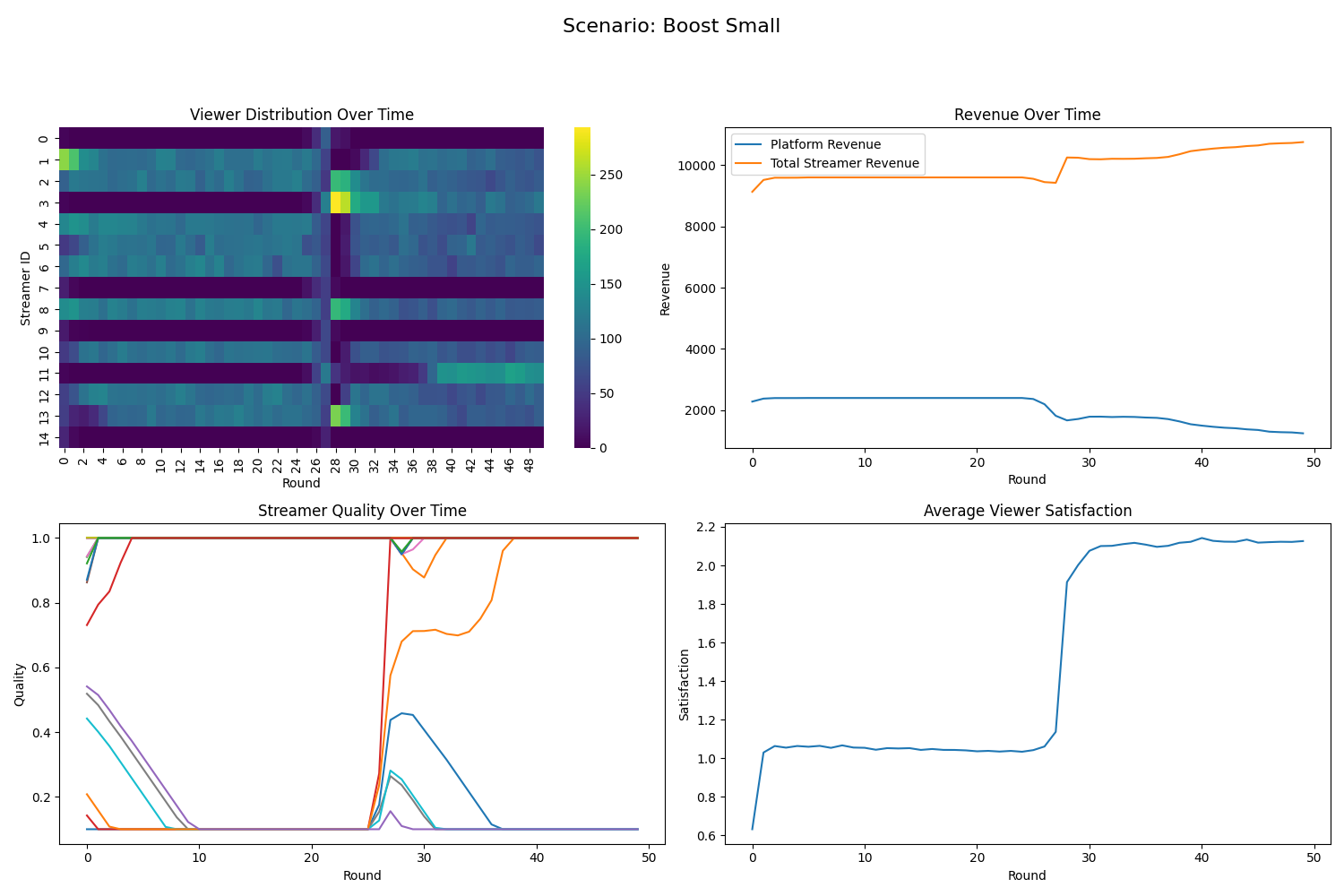}
    \caption{Boost\_Small scenario: Viewer distribution, revenue trends, streamer quality, and average satisfaction over time.}
    \label{fig:figure3}
\end{figure}

\begin{figure}[htbp]
    \centering
    \includegraphics[width=\columnwidth]{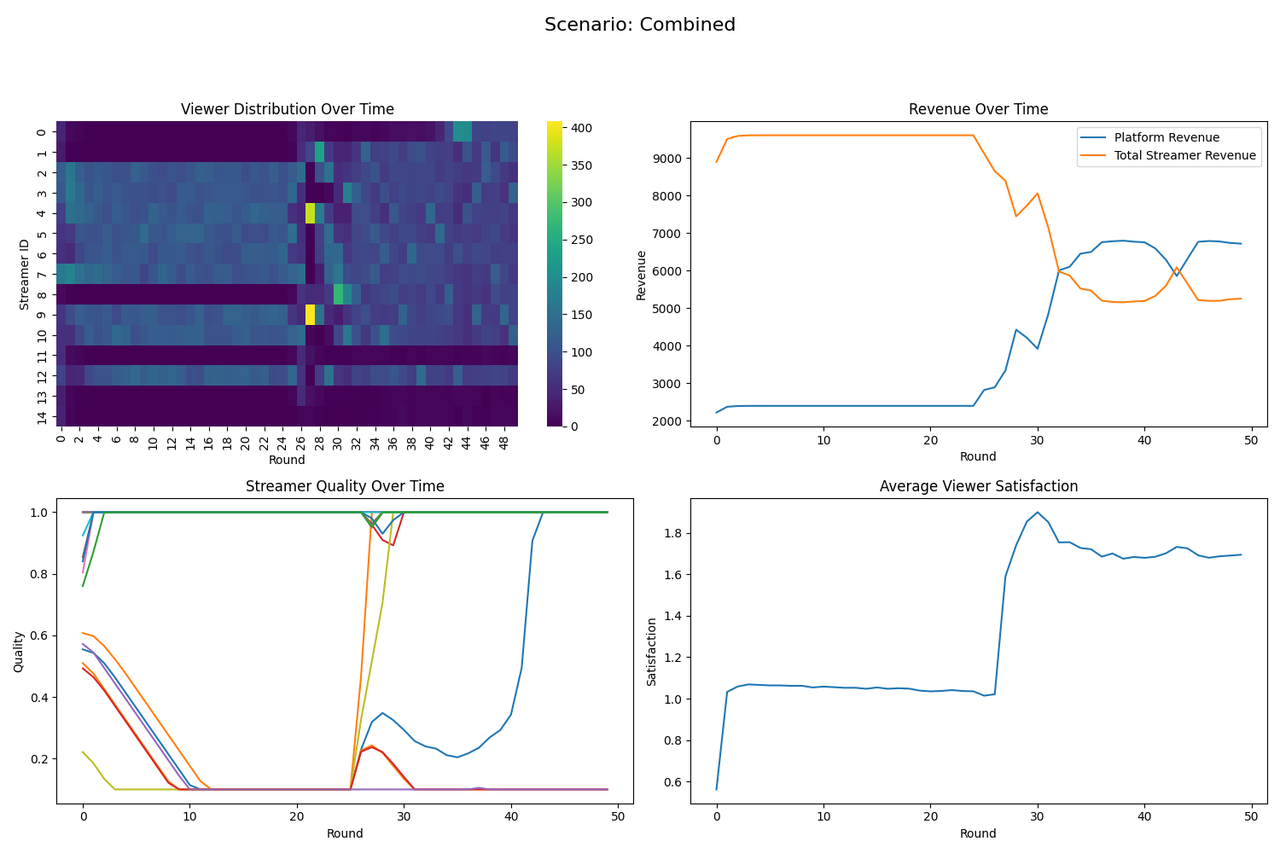}
    \caption{Combined scenario: Viewer distribution, revenue trends, streamer quality, and average satisfaction over time.}
    \label{fig:figure4}
\end{figure}

In this section, we analyze simulation results for four scenarios: \emph{Baseline}, \emph{High\_Tax}, \emph{Boost\_Small}, and \emph{Combined}. Table~\ref{tab:exp_metrics} presents key evaluation metrics, and Figures~\ref{fig:figure2}, \ref{fig:figure3}, and \ref{fig:figure4} show the temporal evolution of viewer distribution, revenue trends, streamer quality, and average viewer satisfaction under different policies.


\paragraph{High\_Tax:}

Imposing a higher commission rate on top streamers (\emph{High\_Tax}) reduces market concentration (Gini coefficient: 0.4816; Top 3 Share: 0.400). Viewer mobility increases to 9.5483, but average viewer satisfaction drops to 0.7966 due to reduced streamer investment. The net quality change of \(- 0.02299\) indicates stagnant quality improvement.

\paragraph{Boost\_Small:}

Granting additional exposure and incentives to lesser - known streamers (\emph{Boost\_Small}) leads to a more balanced distribution (Gini coefficient: 0.3188; Top 3 Share: 0.319; Tail Share: 0.681). Average viewer satisfaction reaches 2.1232, and viewer mobility is 12.1660. However, the net quality improvement is slightly negative (\(- 0.0356\)).

\paragraph{Combined:}

Adopting both policies (\emph{High\_Tax} + \emph{Boost\_Small}) achieves the lowest market concentration (Gini coefficient: 0.2636; Top 3 Share: 0.301). Viewer mobility peaks at 18.8245. Although average viewer satisfaction (1.7533) is lower than in the \emph{Boost\_Small} scenario, it is higher than in the \emph{Baseline} and \emph{High\_Tax} scenarios. The positive quality improvement (0.0966) suggests that resource redistribution promotes diversity and sustained quality investment.

\subsection{Experimental Conclusions}

The experimental results demonstrate that a combined policy approach delivers the best overall outcomes in revenue distribution, market vitality, and content quality. This strategy involves implementing moderate taxes to prevent the excessive concentration of top streamers while simultaneously supporting smaller creators, effectively balancing multiple objectives. In contrast, single policy measures may enhance specific metrics but often lead to unintended negative consequences in other areas.

We recommend adopting a combined policy framework that gradually adjusts tax rates and support mechanisms to strike a balance between efficiency and fairness. It is crucial to monitor how both streamers and viewers respond to these new guidelines. A more equitable revenue distribution supports the platform’s long-term growth and fosters the discovery and development of high-potential streamers. Additionally, increased viewer mobility enhances market dynamism and innovation. Lastly, the observed improvements in content quality indicate that well-calibrated policy interventions can guide the industry toward sustainable and mutually beneficial growth.

\section{Conclusion}

In this study, we constructed a comprehensive theoretical framework complemented by a based model to dissect live-streaming platforms as two-sided markets. Our research delved deep into the emergence of the head effect, wherein a handful of top streamers attract an outsized portion of the audience, a phenomenon propelled by potent network effects and platform-specific policies. By leveraging both static and dynamic models, we systematically explored the intricate interactions among crucial factors, including content quality, pricing strategies, and algorithmic recommendations, and elucidated how these elements jointly shape viewer distribution, streamer revenue, and overall social welfare.

Our experimental evaluation uncovered that a combined policy approach, involving the moderation of top-streamer concentration through taxation and the support of smaller creators, yields the most favorable outcomes in terms of revenue distribution, market vitality, and quality enhancement. This finding provides actionable strategies for platform designers and regulators striving to foster fairness and innovation within digital content markets.

The contributions of this work lie not only in validating the theoretical model through simulation but also in offering a nuanced understanding of the complex dynamics within live-streaming platforms. However, the current simulation framework, despite its ability to capture numerous real-world aspects, has its limitations.

For future research, empirical validation is essential to bridge the gap between the simulated and real-world scenarios. Additionally, expanding the analysis to multi-platform scenarios will enable a more comprehensive view of how competition and cooperation across platforms impact the ecosystem. Considering heterogeneous viewer preferences will add more realism to the model, as viewers' diverse tastes and needs play a significant role in shaping the market. Moreover, exploring more complex revenue models will provide a deeper understanding of the economic mechanisms at work. By pursuing these research directions, we can gain a more profound understanding of the evolving digital economy and, in turn, contribute to the formulation of more effective regulatory policies.

\appendix

\section{Appendix : Experiment Details}
\label{sec:experimental_setup}

\label{sec:experimental_setup}

This section provides a comprehensive overview of our experimental setup designed to evaluate the theoretical model of live streaming platforms through extensive experiments. Our experiment framework, implemented in Python, captures the dynamic interactions between streamers and viewers over multiple rounds. The experiments are structured to assess the performance of the platform under various conditions and policy interventions. Specifically, we:

\begin{itemize}
    \item Define a set of key metrics that quantify platform performance, including market concentration, viewer behavior, and content quality enhancement.
    \item Describe the experiment parameters and initial conditions, detailing the configuration of streamers and viewers, the dynamic parameters governing the model, and the data structures used for tracking experiment outcomes.
    \item Conduct a sensitivity analysis to examine the robustness of the model by varying core parameters—such as the network effect coefficient, base revenue share, number of streamers, and number of viewers—and evaluating their impacts on the key performance metrics.
\end{itemize}

Together, these experimental details provide a robust foundation for understanding the interplay between platform policies and market dynamics, as well as for validating the theoretical predictions of our model.

\subsection{Experiment Metrics and Explanations}

The experimental evaluation relies on several key metrics derived from our experiment outputs. These metrics quantify various aspects of platform dynamics and are defined as follows:

\begin{itemize}
    \item \textbf{Gini Coefficient:} This metric measures the inequality in the distribution of viewers among streamers. It is computed using a custom Gini function, with values ranging from 0 (perfect equality) to 1 (complete inequality). A higher Gini coefficient indicates that a small number of streamers are capturing a disproportionate share of the total viewers.
    
    \item \textbf{Top 3 Share:} Defined as the ratio of the total number of viewers for the top three streamers to the overall viewer count, this metric assesses the degree of concentration in viewer attention. A higher top 3 share suggests a stronger head effect.
    
    \item \textbf{Viewer Mobility:} This metric represents the mean absolute change in viewer counts between consecutive rounds. It quantifies the dynamism in viewer behavior, indicating how frequently viewers switch between streamers.
    
    \item \textbf{Tail Share:} Calculated as the ratio of the total viewers for all streamers, excluding the top three, to the overall viewer count, the tail share reflects the diversity of viewership among less popular streamers.
    
    \item \textbf{Average Satisfaction:} The average satisfaction score of viewers in the final experiment round is used to gauge overall viewer experience. This score is derived from the computed utility values, incorporating factors such as content quality, network effects, and price sensitivity.
    
    \item \textbf{Quality Improvement:} This metric measures the average change in streamers’ content quality from the initial to the final round, capturing the extent of investment in content enhancement over time.
\end{itemize}

Together, these metrics provide a comprehensive assessment of the platform’s performance, offering insights into market concentration, audience behavior, and the effectiveness of quality improvement strategies under different policy interventions.

\subsection{ Parameters and Initial Conditions}

Our experiment framework is implemented in Python to model interactions between streamers and viewers over multiple rounds. The following subsections describe the key components of our experimental setup.

\paragraph{Data Structures:} We define two primary configurations using Python’s \texttt{dataclass}:

\begin{itemize}
    \item \textbf{StreamerConfig} includes:
    \begin{itemize}
        \item \textit{initial\_quality:} Sampled from a normal distribution with mean 0.5 and standard deviation 0.2, then clipped to the interval [0.1, 0.9].
        \item \textit{current\_quality:} Initially set equal to \textit{initial\_quality}.
        \item \textit{cost\_coefficient:} A uniformly distributed random value between 0.1 and 0.3.
        \item \textit{revenue\_share:} Set to the base revenue share (default 0.2).
        \item \textit{exposure\_boost:} Initialized to 1.0.
        \item \textit{followers:} Initialized to 0.
    \end{itemize}
    \item \textbf{ViewerConfig} includes:
    \begin{itemize}
        \item \textit{interaction\_willingness:} Uniformly sampled between 0.2 and 0.8.
        \item \textit{price\_sensitivity:} Uniformly sampled between 0.3 and 0.7.
        \item \textit{quality\_sensitivity:} Uniformly sampled between 0.4 and 0.8.
        \item \textit{network\_effect\_sensitivity:} Uniformly sampled between 0.1 and 0.4.
        \item \textit{preferred\_content\_type:} A random integer between 0 and 2.
        \item \textit{loyalty:} Uniformly sampled between 0.3 and 0.7.
    \end{itemize}
\end{itemize}

\paragraph{Platform Initialization:} The \texttt{LiveStreamingPlatform} class is instantiated with the following default parameters:
\begin{itemize}
    \item \textbf{Number of Streamers:} 15.
    \item \textbf{Number of Viewers:} 1000.
    \item \textbf{Number of Rounds:} 50.
    \item \textbf{Base Revenue Share:} 0.2.
    \item \textbf{Network Effect Parameter (\(\beta\)):} 0.15.
    \item \textbf{Quality Decay Rate:} 0.01.
    \item \textbf{Random Effect Scale:} 0.2.
\end{itemize}
These parameters govern the dynamic interactions in the experiment, including viewer behavior, content quality evolution, and revenue generation.

\paragraph{Tracking Variables:} Throughout the experiment, several key metrics are recorded in each round:
\begin{itemize}
    \item \textbf{Viewer Distribution:} The number of viewers per streamer.
    \item \textbf{Streamer Revenues:} The revenue earned by each streamer.
    \item \textbf{Platform Revenue:} The aggregated revenue collected by the platform.
    \item \textbf{Quality History:} The evolution of each streamer's content quality.
    \item \textbf{Viewer Satisfaction:} The satisfaction level of each viewer.
\end{itemize}

\paragraph{Policy Interventions:} Our framework also supports the implementation of various intervention policies (e.g., imposing a high commission rate for top streamers or boosting exposure for smaller streamers) at designated rounds. This functionality enables us to assess the impact of such policies on traffic allocation, market concentration, and overall social welfare.

This comprehensive setup ensures that our experiment accurately captures the dynamics of live streaming platforms, providing a robust foundation for experimental evaluation.

\subsection{Sensitivity Analysis Details}

\begin{figure}[htbp]
    \centering
    \includegraphics[width=\columnwidth]{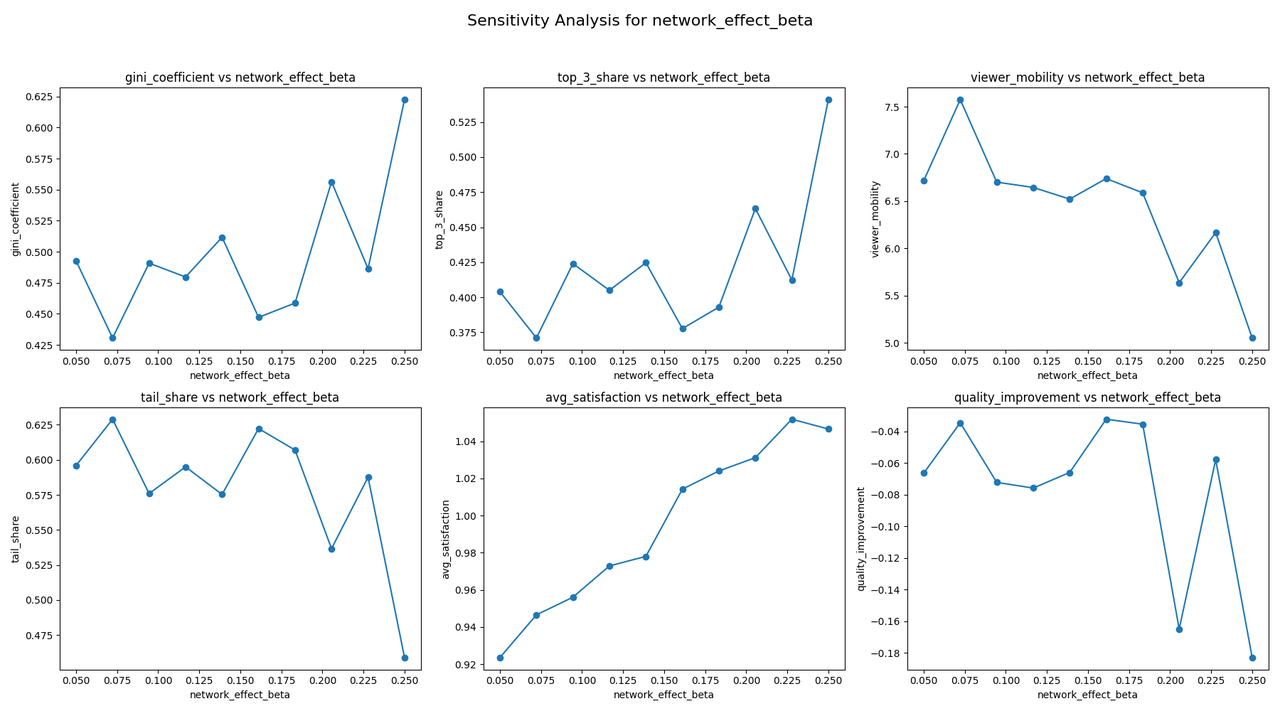}
    \caption{Sensitivity analysis for \texttt{network\_effect\_beta}, showing the impact on key metrics.}
    \label{fig:figure5}
\end{figure}

\begin{figure}[htbp]
    \centering
    \includegraphics[width=\columnwidth]{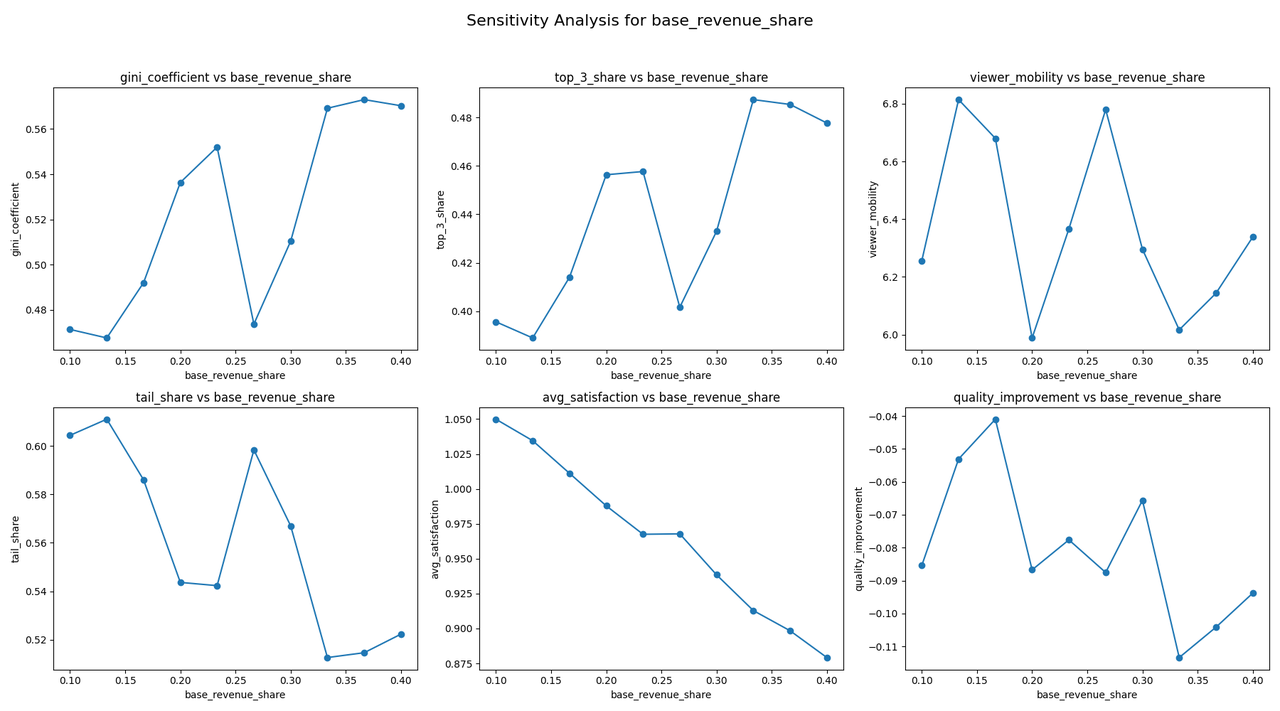}
    \caption{Sensitivity analysis for \texttt{base\_revenue\_share}, illustrating its effect on key performance metrics.}
    \label{fig:figure6}
\end{figure}

\begin{figure}[htbp]
    \centering
    \includegraphics[width=\columnwidth]{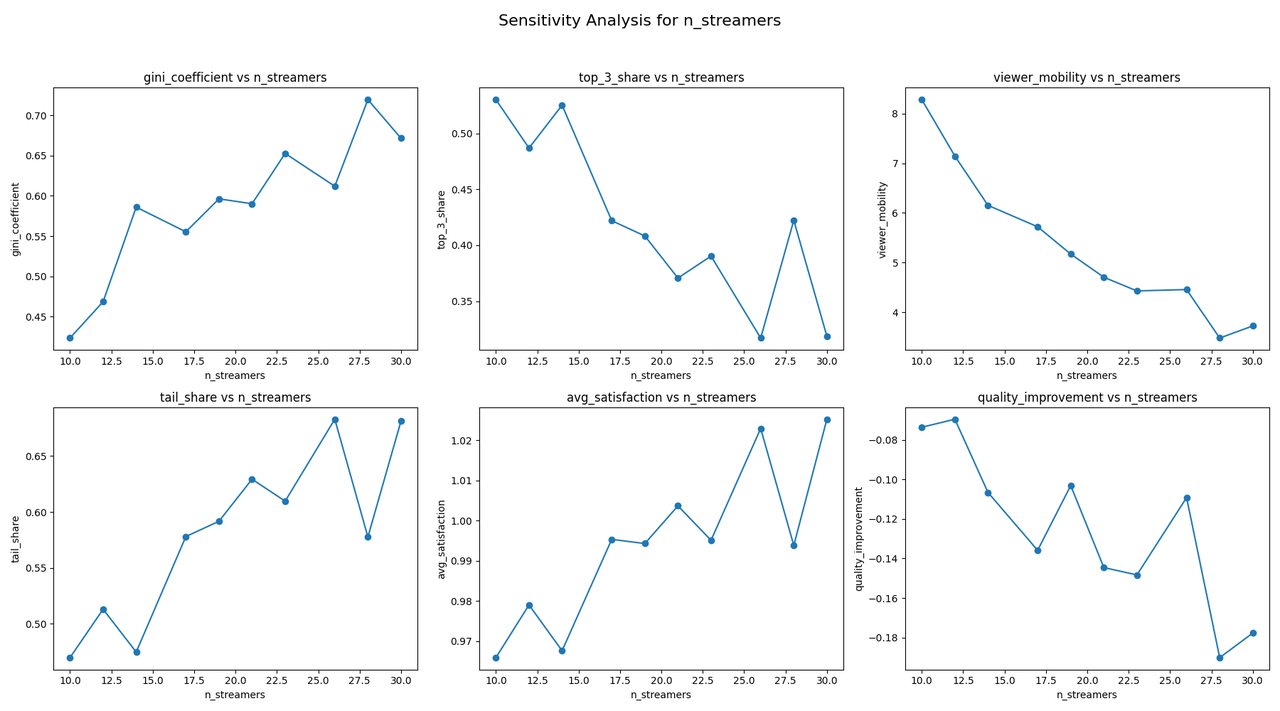}
    \caption{Sensitivity analysis for \texttt{n\_streamers}, demonstrating the impact on concentration, satisfaction, and quality dynamics.}
    \label{fig:figure7}
\end{figure}

\begin{figure}[htbp]
    \centering
    \includegraphics[width=\columnwidth]{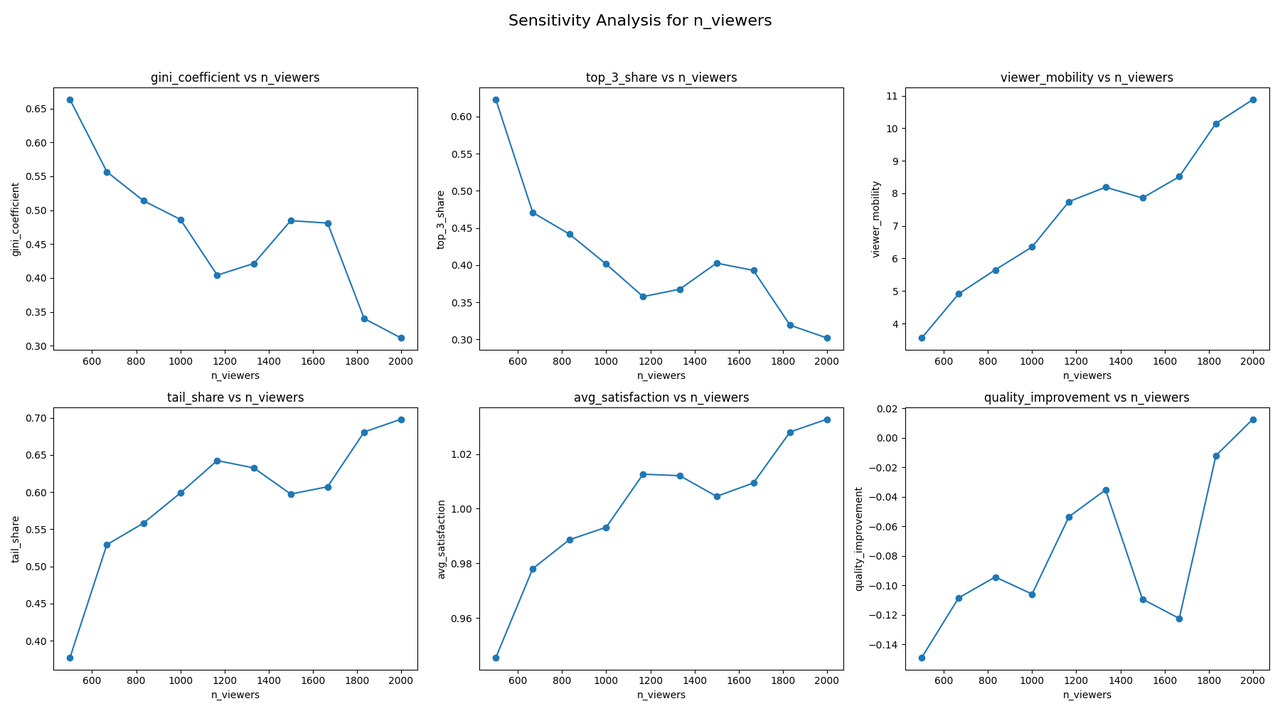}
    \caption{Sensitivity analysis for \texttt{n\_viewers}, highlighting the influence of audience size on distribution, satisfaction, and quality enhancement.}
    \label{fig:figure8}
\end{figure}

In this subsection, we conduct a comprehensive sensitivity analysis to illustrate how our experiment responds to variations in key parameters. We focus on four main parameters: (i) the network effect coefficient (\texttt{network\_effect\_beta}), (ii) the platform base revenue share (\texttt{base\_revenue\_share}), (iii) the number of streamers (\texttt{n\_streamers}), and (iv) the number of viewers (\texttt{n\_viewers}). Figures~\ref{fig:figure5}--\ref{fig:figure8} depict the corresponding impacts on metrics such as the Gini coefficient, Top 3 Share, viewer mobility, tail share, average satisfaction, and quality improvement.

\paragraph{Network Effect Coefficient (\texttt{network\_effect\_beta}):}  
Increasing the network effect coefficient from 0.05 to 0.25 generally results in:
\begin{itemize}
    \item \textbf{Increased Market Concentration:} Both the Gini coefficient and Top 3 Share rise, indicating that viewers tend to cluster around a few popular streamers, which exacerbates the head effect.
    \item \textbf{Higher Viewer Satisfaction but Lower Mobility:} Although a stronger network effect boosts viewer satisfaction as audiences follow popular channels, the reduced mobility suggests a more static yet polarized audience distribution.
    \item \textbf{Reduced Quality Improvement:} With a more pronounced head effect, mid-to-small streamers face challenges in investing in quality, leading to a lower overall improvement.
\end{itemize}

\paragraph{Platform Base Revenue Share (\texttt{base\_revenue\_share}):}  
Varying the base revenue share from 0.10 to 0.40 reveals that:
\begin{itemize}
    \item \textbf{Elevated Market Inequality:} Higher commission rates lead to an increase in both the Gini coefficient and Top 3 Share, suggesting that top streamers consolidate their positions while smaller streamers lose ground.
    \item \textbf{Declining Viewer Satisfaction:} As the platform's share increases, overall viewer satisfaction tends to decline, likely due to reduced streamer investment in content quality or increased monetization pressures.
    \item \textbf{Diminished Quality Improvement:} Reduced net revenue limits the resources available for streamers to enhance content quality.
\end{itemize}

\paragraph{Number of Streamers (\texttt{n\_streamers}):}  
Increasing the number of streamers from 10 to 30 leads to:
\begin{itemize}
    \item \textbf{Increased Market Concentration:} Despite a larger pool of streamers, the Gini coefficient increases as viewers still tend to concentrate on a few standout channels.
    \item \textbf{Reduced Viewer Mobility with Slightly Higher Satisfaction:} A larger streamer pool reduces the frequency of viewer switching, while niche content offerings contribute to a modest improvement in overall satisfaction.
    \item \textbf{Worsening Quality Improvement:} Intensified competition makes it challenging for new entrants to maintain investments in content quality, resulting in a downward trend in quality improvement.
\end{itemize}

\paragraph{Number of Viewers (\texttt{n\_viewers}):}  
Expanding the viewer base from 500 to 2000 produces the following trends:
\begin{itemize}
    \item \textbf{More Balanced Traffic Distribution:} The Gini coefficient and Top 3 Share decrease, while the tail share increases, indicating that a larger audience promotes greater diversity in viewership.
    \item \textbf{Higher Viewer Mobility and Satisfaction:} An increased number of viewers leads to more dynamic audience behavior and a notable rise in average satisfaction, as diverse preferences are better accommodated.
    \item \textbf{Enhanced Quality Investment:} With a larger audience, streamers have stronger incentives to invest in content quality, leading to an improvement from negative to near-zero or slightly positive quality gains.
\end{itemize}

\paragraph{Overall Insights:}  
In summary, the sensitivity analyses reveal that:
\begin{itemize}
    \item \textbf{Network Effects \& Platform Commission:} Stronger network effects, while improving viewer satisfaction, also intensify market concentration. Similarly, higher platform commissions exacerbate inequality and reduce the capacity for quality improvement.
    \item \textbf{Population Sizes:} Although increasing the number of streamers can enhance long-tail content diversity, it also heightens competition, leading to uneven success. Conversely, expanding the viewer base tends to yield a more equitable distribution of traffic, higher satisfaction, and better incentives for quality investments.
\end{itemize}
These results underscore the need for a balanced approach in setting platform policies to support both competitive dynamics and sustainable growth.

\section{Appendix : Unified Notation and Definitions}

This appendix provides detailed theoretical proofs and derivations that support the claims and results presented in the main text. It is organized into two main sections,In this appendix, we use the following notation which is consistent with the main text:
\begin{itemize}
    \item \( \beta \): The network effect parameter.
    \item \( n_i \): The number of viewers for streamer \(i\).
    \item \( q_i \): The content quality of streamer \(i\).
    \item \( \Delta q = q_{i^*} - q_j \): The difference in content quality between the top streamer \(i^*\) and any other streamer \(j\).
    \item \( M \): The total number of viewers.
    \item \( N \): The total number of streamers.
    \item \( \phi \): A sensitivity factor defined as the marginal sensitivity of the expected utility (or equivalently, the choice probability) with respect to the traffic allocation parameter \(\theta_i\). In particular, we assume that
    \[
    \frac{\partial E[U_j]}{\partial \theta_i} = \frac{P_i}{\phi},
    \]
    and, using the multinomial logit model property,
    \[
    \frac{\partial n_i}{\partial \theta_i} = M \frac{\partial P_i}{\partial \theta_i} = M P_i (1 - P_i) \phi.
    \]
\end{itemize}

All additional symbols such as \( \tau \) (platform commission rate), \( R \) (revenue per viewer), and parameters in the cost function \( c(q_i) \) are defined in the main text.

\section{Appendix : Detailed Theoretical Proofs and Derivations in Static Analysis}

\subsection{Proof of Theorem 1}
\begin{theorem}[Critical Network Effect]
There exists a critical value \( \beta^* > 0 \) such that if \( \beta > \beta^* \), the market equilibrium results in one streamer capturing almost all viewers.
\end{theorem}

\textbf{Proof:}  
We aim to show that when the network effect parameter \( \beta \) exceeds a certain critical value \( \beta^* \), the equilibrium viewer distribution becomes highly concentrated, with one streamer (denoted \( i^* \)) capturing nearly all viewers.

\noindent\textbf{Step 1.} Consider the ratio of the probabilities that a viewer chooses streamer \( i^* \) versus another streamer \( j \). Under the multinomial logit model, we have
\begin{equation}\label{eq:ratio1}
\frac{P_{i^*}}{P_j} = \exp\Bigl( \alpha\, \Delta q + \beta (n_{i^*} - n_j) \Bigr),
\end{equation}
where we define 
\[
\Delta q = q_{i^*} - q_j,
\]
which represents the difference in content quality between the top streamer \( i^* \) and another streamer \( j \).

\noindent\textbf{Step 2.} Since the total number of viewers is \( M \), we have:
\begin{equation}\label{eq:total_viewers}
n_{i^*} + (N - 1)n_j = M.
\end{equation}
Thus, expressing \( n_j \) in terms of \( n_{i^*} \):
\begin{equation}\label{eq:nj}
n_j = \frac{M - n_{i^*}}{N - 1}.
\end{equation}

\noindent\textbf{Step 3.} Substitute \eqref{eq:nj} into \eqref{eq:ratio1}:
\begin{equation}\label{eq:ratio2}
\frac{P_{i^*}}{P_j} = \exp\left( \alpha\, \Delta q + \beta \left( n_{i^*} - \frac{M - n_{i^*}}{N - 1} \right) \right).
\end{equation}

\noindent\textbf{Step 4.} Simplify the term in the exponent:
\begin{align}
n_{i^*} - \frac{M - n_{i^*}}{N - 1} &= \frac{(N-1)n_{i^*} - (M - n_{i^*})}{N-1} \nonumber\\
&= \frac{N n_{i^*} - M}{N - 1}. \label{eq:simplify}
\end{align}
Assuming \( N \) is large (i.e., \(N-1 \approx N\)), we approximate \eqref{eq:simplify} as
\[
\approx n_{i^*} - \frac{M}{N}.
\]
Thus, \eqref{eq:ratio2} becomes
\begin{equation}\label{eq:ratio3}
\frac{P_{i^*}}{P_j} \approx \exp\Bigl( \alpha\, \Delta q + \beta \left( n_{i^*} - \frac{M}{N} \right) \Bigr).
\end{equation}

\noindent\textbf{Step 5.} Since the expected number of viewers is given by \( n_i = M P_i \), we have:
\[
\frac{n_{i^*}}{n_j} = \frac{P_{i^*}}{P_j} \approx \exp\Bigl( \alpha\, \Delta q + \beta (n_{i^*} - n_j) \Bigr).
\]
Substitute again \( n_j = \frac{M - n_{i^*}}{N - 1} \) into the above ratio:
\begin{equation}\label{eq:ratio4}
\frac{n_{i^*}}{\frac{M - n_{i^*}}{N - 1}} = \exp\Bigl( \alpha\, \Delta q + \beta \left( n_{i^*} - \frac{M - n_{i^*}}{N - 1} \right) \Bigr).
\end{equation}

\noindent\textbf{Step 6.} The left-hand side of \eqref{eq:ratio4} simplifies to
\[
(N - 1) \frac{n_{i^*}}{M - n_{i^*}}.
\]
Again, for large \( N \) (i.e., \(N-1 \approx N\)), we have:
\[
N \cdot \frac{n_{i^*}}{M - n_{i^*}} \approx \exp\Bigl( \alpha\, \Delta q + \beta n_{i^*} \Bigr),
\]
where we have approximated \(\frac{M - n_{i^*}}{N} \approx 0\) when \( n_{i^*} \) is very close to \( M \) (i.e., in the head effect scenario).

\noindent\textbf{Step 7.} Taking natural logarithms on both sides yields:
\begin{equation}\label{eq:log_form}
\ln N + \ln\left( \frac{n_{i^*}}{M - n_{i^*}} \right) = \alpha\, \Delta q + \beta n_{i^*}.
\end{equation}

Assume that in the extreme head effect, \( n_{i^*} \approx M - \epsilon \) with \( \epsilon \ll M \), then:
\[
\frac{n_{i^*}}{M - n_{i^*}} \approx \frac{M - \epsilon}{\epsilon} \approx \frac{M}{\epsilon}.
\]
Thus, \eqref{eq:log_form} becomes:
\begin{equation}\label{eq:final_log}
\ln N + \ln M - \ln \epsilon = \alpha\, \Delta q + \beta (M - \epsilon).
\end{equation}

\noindent\textbf{Step 8.} Note that for large \( M \) and \( N \), the terms \( \ln M \) and \( \ln N \) grow logarithmically, while \( \beta M \) grows linearly. In order for \eqref{eq:final_log} to hold, it is necessary that \( \beta M \) is sufficiently large. Hence, there exists a critical value \( \beta^* \) such that if \( \beta > \beta^* \), then \( n_{i^*} \to M \).

Therefore, the top streamer \( i^* \) captures almost all viewers when \( \beta > \beta^* \). \hfill\(\Box\)

\subsection{Proof of Theorem 2}
\begin{theorem}[Uniqueness of Steady-State]
Under the given parameters and assumptions, there exists a unique steady-state solution \( (n_i^*, q_i^*) \).
\end{theorem}

\textbf{Proof:}  
We consider the static system in which streamers choose their content quality \( q_i \) to maximize profit and viewers choose streamers according to the multinomial logit model.

\noindent\textbf{Existence:}  
\begin{enumerate}
    \item The set of feasible viewer numbers \( n_i \) is bounded between 0 and \( M \), and the content quality \( q_i \) is bounded below (since producing zero quality is not viable) and above (due to prohibitive costs at very high quality).
    \item The best-response functions derived from the streamers' optimization problems are continuous mappings from a compact set to itself.
    \item By Brouwer's Fixed Point Theorem (see, e.g., \cite{Browder1968}), there exists at least one fixed point \( (n_i^*, q_i^*) \).
\end{enumerate}

\noindent\textbf{Uniqueness:}  
\begin{enumerate}
    \item The cost function \( c(q_i) \) is strictly convex (i.e., \( c''(q_i) > 0 \)), implying that the profit function is strictly concave in \( q_i \).
    \item The viewer choice probabilities \( P_i \) are strictly increasing in \( \alpha_i q_i \) and \( n_i \).
    \item This strict monotonicity ensures that the mapping from \( q_i \) to \( n_i \) and vice versa is one-to-one.
\end{enumerate}

Thus, the steady-state solution \( (n_i^*, q_i^*) \) is unique. \hfill\(\Box\)

\subsection{Proof of Theorem 3}
\begin{theorem}[Local Asymptotic Stability]
The steady state \( (n_i^*, q_i^*) \) is locally asymptotically stable if all eigenvalues of the Jacobian matrix \( J \) have negative real parts.
\end{theorem}

\textbf{Proof:}  
We analyze the system by linearizing the dynamic equations around the steady state.

\noindent\textbf{Step 1.} Define the perturbations:
\[
\delta n_i = n_i - n_i^*, \quad \delta q_i = q_i - q_i^*.
\]
Then, the linearized system can be written as:
\begin{equation}\label{eq:linearized}
\frac{d}{dt} \begin{pmatrix} \delta n_i \\ \delta q_i \end{pmatrix} = J \begin{pmatrix} \delta n_i \\ \delta q_i \end{pmatrix},
\end{equation}
where \( J \) is the Jacobian matrix evaluated at \( (n_i^*, q_i^*) \).

\noindent\textbf{Step 2.} The Jacobian matrix \( J \) is given by:
\begin{equation}\label{eq:Jacobian}
J = \begin{pmatrix}
\frac{\partial \dot{n}_i}{\partial n_i} & \frac{\partial \dot{n}_i}{\partial q_i} \\[1mm]
\frac{\partial \dot{q}_i}{\partial n_i} & \frac{\partial \dot{q}_i}{\partial q_i}
\end{pmatrix}.
\end{equation}

The partial derivatives are computed as follows:
\begin{enumerate}
    \item \( \displaystyle \frac{\partial \dot{n}_i}{\partial n_i} = \gamma \left( M \frac{\partial P_i}{\partial n_i} - 1 \right). \)
    \item \( \displaystyle \frac{\partial \dot{n}_i}{\partial q_i} = \gamma M \frac{\partial P_i}{\partial q_i}. \)
    \item For \( \dot{q}_i \), from the first-order condition of the streamer's optimization:
    \[
    c'(q_i) = (1-\tau) R M P_i (1-P_i) \alpha_i,
    \]
    we differentiate with respect to \( n_i \) and \( q_i \) (detailed intermediate steps are omitted here for brevity; see standard derivations in \cite{Basar1999}).
\end{enumerate}

\noindent\textbf{Step 3.} By applying the Routh-Hurwitz criterion (or a similar stability result), if all eigenvalues of \( J \) satisfy \( \text{Re}(\lambda) < 0 \), then the steady state is locally asymptotically stable.

Therefore, the steady state is locally asymptotically stable. \hfill\(\Box\)


\section{Appendix : Detailed Theoretical Proofs and Derivations in Dynamic Analysis}

\subsection{Existence and Uniqueness of Steady-State Solutions under the Dynamic Model}

\begin{theorem}[Dynamic Steady-State]
Under the stated assumptions, the dynamic system has a unique steady-state solution \( (n_i^*, q_i^*) \), where \( n_i^* \) is the steady-state number of viewers and \( q_i^* \) is the steady-state content quality for streamer \( i \).
\end{theorem}

\textbf{Proof:}  
\textbf{Assumptions:}
\begin{itemize}
    \item \textbf{Continuity and Boundedness:} The functions in the dynamic equations are continuous and operate on compact domains.
    \item \textbf{Strict Convexity:} The cost function \( c(q_i) \) is strictly convex, i.e., \( c''(q_i) > 0 \).
    \item \textbf{Monotonicity:} The response functions are monotonic with respect to their arguments.
    \item \textbf{Positive Parameters:} All parameters such as \( \alpha_i \), \( \beta \), \( \gamma \), \( M \), and \( R \) are positive.
\end{itemize}

\medskip

\noindent\textbf{Dynamic Equations:}
\begin{itemize}
    \item \emph{Viewer Dynamics:}
    \[
    \frac{d n_i(t)}{d t} = \gamma \left[ M P_i(t) - n_i(t) \right],
    \]
    where
    \[
    P_i(t) = \frac{\exp\bigl(V_i(t)\bigr)}{\sum_{k=1}^N \exp\bigl(V_k(t)\bigr)}
    \]
    and
    \[
    V_i(t) = \alpha_i q_i(t) - p_i + \beta n_i(t).
    \]
    
    \item \emph{Streamer Optimization:}  
    Streamers choose \( q_i(t) \) to maximize profit, with the revenue function given by
    \[
    V_i(t) = (1-\tau) R\, n_i(t) - c\bigl(q_i(t)\bigr).
    \]
    The first-order condition (FOC) is:
    \[
    c'\bigl(q_i(t)\bigr) = (1-\tau) R \frac{\partial n_i(t)}{\partial q_i(t)}.
    \]
\end{itemize}

\medskip

\noindent\textbf{Proof Steps:}
\begin{enumerate}
    \item \emph{Existence:} Since the functions are continuous and the domain is compact (due to bounded \( n_i(t) \) and \( q_i(t) \)), by Brouwer's Fixed Point Theorem there exists at least one fixed point \( (n_i^*, q_i^*) \).
    \item \emph{Uniqueness:} The strict convexity of \( c(q_i) \) and the monotonicity of the best-response functions ensure that the mapping from \( q_i \) to \( n_i \) is one-to-one, which guarantees a unique steady-state solution.
\end{enumerate}

Thus, the steady-state solution \( (n_i^*, q_i^*) \) is unique. \hfill\(\Box\)

\bigskip

\subsection{Proof of Path Dependence under the Dynamic Model}
\begin{theorem}[Path Dependence]
In the dynamic model with network effects, small differences in the initial viewer numbers \( n_i(0) \) can lead to significantly different long-term outcomes \( n_i^* \) due to path dependence, especially when \( \beta \) is sufficiently large.
\end{theorem}

\textbf{Proof:}  
Assume that all streamers have identical parameters except for their initial viewer numbers, i.e., \( \alpha_i = \alpha \), \( q_i(0) = q_0 \), and \( p_i = p \).

\medskip

\noindent\textbf{Step 1.} The viewer dynamics are given by:
\[
\frac{d n_i(t)}{d t} = \gamma \left[ M P_i(t) - n_i(t) \right].
\]

\noindent\textbf{Step 2.} Let \( \Delta n(t) = n_i(t) - n_j(t) \) for two streamers \( i \) and \( j \) with \( n_i(0) > n_j(0) \). Then, the differential equation for the difference is:
\[
\frac{d \Delta n(t)}{d t} = \gamma \Bigl[ M (P_i(t) - P_j(t)) - \Delta n(t) \Bigr].
\]

\noindent\textbf{Step 3.} Since \( P_i(t) > P_j(t) \) (due to the higher initial \( n_i(0) \) and the network effect), the term \( M (P_i(t) - P_j(t)) \) is positive and amplifies the difference \( \Delta n(t) \) over time.

\medskip

Thus, as \( t \to \infty \), even a small initial difference \( \Delta n(0) \) becomes significantly larger, demonstrating path dependence. \hfill\(\Box\)

\bigskip

\subsection{Proof of the Head Effect under the Dynamic Model}
\begin{theorem}[Head Effect in the Dynamic Model]
For sufficiently large \( \beta \) (i.e., strong network effects), the dynamic model exhibits the head effect, wherein one or a few streamers capture the majority of viewers, leading to high market concentration.
\end{theorem}

\textbf{Proof:}  
Assume all streamers have identical intrinsic characteristics (\( \alpha_i = \alpha \), \( p_i = p \), \( q_i(0) = q_0 \)) except for small random variations in initial viewer numbers.

\medskip

\noindent\textbf{Step 1.} The attractiveness function \( V_i(t) \) contains the term \( \beta n_i(t) \), which becomes dominant when \( \beta \) is large.
  
\noindent\textbf{Step 2.} Small initial differences in \( n_i(0) \) lead to differences in \( V_i(0) \) and, by the properties of the multinomial logit model, to differences in \( P_i(t) \). This creates a positive feedback loop, amplifying the advantage of the initially better-performing streamers.

\medskip

\noindent\textbf{Step 3.} Define the market share as
\[
s_i(t) = \frac{n_i(t)}{M},
\]
and the Herfindahl-Hirschman Index (HHI) as
\[
\text{HHI}(t) = \sum_{i=1}^N s_i(t)^2.
\]
As \( t \to \infty \), the dominant streamer will have \( s_i(t) \to 1 \) and the others \( s_j(t) \to 0 \), so that \( \text{HHI}(t) \to 1 \).

\medskip

Therefore, under strong network effects, the head effect emerges. \hfill\(\Box\)

\bigskip

\subsection{Welfare Impact and Derivation of Optimal Traffic Allocation Policy}
\begin{proposition}[Welfare Impact under the Head Effect]
The head effect can lead to a decrease in total social welfare due to reduced consumer surplus (from a lack of diversity) and a potential decrease in producer surplus as many streamers exit the market.
\end{proposition}

\textbf{Proof:}
\begin{enumerate}
    \item \textbf{Consumer Surplus:} Under the head effect, assume \( n_{i^*} \approx M \) for the dominant streamer and \( n_j \approx 0 \) for all \( j \neq i^* \). Then, consumer surplus is approximately
    \[
    CS \approx M \Bigl( \alpha q_{i^*} - p + \beta M \Bigr).
    \]
    Although the network effect term \( \beta M \) increases utility, the reduction in content variety may lower overall consumer surplus.
    
    \item \textbf{Producer Surplus:} The dominant streamer earns
    \[
    PS \approx (1-\tau) R M - c(q_{i^*}),
    \]
    while other streamers earn negligible profits, reducing overall producer surplus.
    
    \item \textbf{Platform Profit:} Although the platform earns \( \Pi = \tau R M \) in the short term, long-term profit may decline if reduced diversity leads to lower viewer engagement.
    
    \item \textbf{Social Welfare:} Comparing total welfare (i.e., \( CS + PS + \Pi \)) with a scenario without the head effect shows that excessive concentration can harm overall welfare.
\end{enumerate}
Thus, the proposition is established. \hfill\(\Box\)

\bigskip

\subsection{Derivation of Optimal Traffic Allocation Policy}
\textbf{Objective:} Maximize total social welfare 
\[
W = CS + PS + \Pi
\]
with respect to the traffic allocation parameters \( \{ \theta_i \} \), subject to
\[
\theta_i \ge 0, \quad \sum_{i=1}^N \theta_i = 1.
\]

\medskip

\noindent\textbf{Lagrangian Formulation:}  
Define the Lagrangian:
\begin{equation}\label{eq:Lagrangian}
\mathcal{L} = W - \lambda \left( \sum_{i=1}^N \theta_i - 1 \right) - \sum_{i=1}^N \mu_i \theta_i,
\end{equation}
where \( \lambda \) is the multiplier for the equality constraint and \( \mu_i \ge 0 \) for the non-negativity constraints.

\medskip

\noindent\textbf{First-Order Conditions:}  
For each \( \theta_i \),
\begin{equation}\label{eq:FOC}
\frac{\partial \mathcal{L}}{\partial \theta_i} = \frac{\partial W}{\partial \theta_i} - \lambda - \mu_i = 0,
\end{equation}
with complementary slackness \( \mu_i \theta_i = 0 \).

\medskip

\noindent\textbf{Derivatives of Social Welfare:}
\begin{enumerate}
    \item \emph{Consumer Surplus Derivative:}  
    We assume
    \[
    \frac{\partial CS}{\partial \theta_i} = M \frac{\partial E[U_j]}{\partial \theta_i} = M \frac{P_i}{\phi}.
    \]
    \item \emph{Producer Surplus Derivative:}  
    \[
    \frac{\partial PS}{\partial \theta_i} = (1-\tau) R \frac{\partial n_i}{\partial \theta_i}.
    \]
    \item \emph{Platform Profit Derivative:}  
    \[
    \frac{\partial \Pi}{\partial \theta_i} = \tau R \frac{\partial n_i}{\partial \theta_i}.
    \]
    \item \emph{Total Derivative of Viewer Numbers:}  
    By the multinomial logit property and incorporating the sensitivity factor \( \phi \),
    \[
    \frac{\partial n_i}{\partial \theta_i} = M \frac{\partial P_i}{\partial \theta_i} = M P_i (1-P_i) \phi.
    \]
\end{enumerate}

\medskip

\noindent\textbf{Simplifying the First-Order Condition:}  
Combining the above, the first-order condition becomes:
\begin{equation}\label{eq:FOC_simplified}
M \frac{P_i}{\phi} + \Bigl[(1-\tau)R + \tau R\Bigr] M P_i (1-P_i) \phi = \lambda.
\end{equation}
Since \((1-\tau)R + \tau R = R\), this simplifies to:
\begin{equation}\label{eq:FOC_final}
\frac{P_i}{\phi} + R M P_i (1-P_i) \phi = \lambda.
\end{equation}

\medskip

\noindent\textbf{Case Analysis:}
\begin{enumerate}
    \item \emph{Interior Solution} (\( \theta_i > 0 \)): Here, \( \mu_i = 0 \) and \eqref{eq:FOC_final} holds.
    \item \emph{Corner Solution} (\( \theta_i = 0 \)): Then, complementary slackness requires
    \[
    M \left( \frac{P_i}{\phi} + R P_i (1-P_i) \phi M \right) - \lambda \le 0.
    \]
\end{enumerate}

\medskip

Thus, the optimal \( \theta_i \) depends on \( P_i \), \( R \), \( M \), and \( \phi \). To maximize overall social welfare, the platform should allocate more traffic to streamers where the marginal increase in social welfare per unit change in \( \theta_i \) is highest. \hfill\(\Box\)

\bigskip

\subsection{Dynamic Optimization Problem}
In the dynamic context, the platform seeks to maximize the discounted sum of welfare over time:
\[
\max_{\{ \theta_i(t) \}} \int_0^\infty e^{-\rho t} W(t) \, dt,
\]
subject to the dynamic equations for \( n_i(t) \) and \( q_i(t) \) and the constraint
\[
\theta_i(t) \ge 0, \quad \sum_{i=1}^N \theta_i(t) = 1,
\]
where \( \rho \) is the discount rate.

\medskip

\noindent\textbf{Hamiltonian Function:}  
Define the current-value Hamiltonian:
\begin{equation}\label{eq:Hamiltonian}
\mathcal{H} = W(t) + \sum_{i=1}^N \lambda_i(t) \left( \gamma \Bigl[ M P_i(t) - n_i(t) \Bigr] \right),
\end{equation}
where \( \lambda_i(t) \) are the costate variables associated with the state equations for \( n_i(t) \).

\medskip

\noindent\textbf{First-Order Conditions:}
\begin{enumerate}
    \item For the control variable \( \theta_i(t) \):
    \[
    \frac{\partial \mathcal{H}}{\partial \theta_i} = 0.
    \]
    \item The costate equations for \( \lambda_i(t) \) are given by:
    \[
    \dot{\lambda}_i(t) = \rho \lambda_i(t) - \frac{\partial \mathcal{H}}{\partial n_i}.
    \]
\end{enumerate}

\medskip

Solving these conditions along with the state equations and boundary conditions yields the optimal paths \( \theta_i(t) \), \( n_i(t) \), and \( \lambda_i(t) \). In many cases, analytical solutions are intractable, and numerical methods or approximations are required.


\bibliographystyle{ACM-Reference-Format}
\bibliography{sample-base}


\end{document}